\documentclass[aps,prl,twocolumn,showpacs,superscriptaddress,groupedaddress,fleqn,reprint]{revtex4}  
\usepackage{graphicx}  
\usepackage{dcolumn}   
\usepackage{bm}        
\usepackage{amssymb}   
\usepackage{color}

\newcommand{\mnras}{Mon. Not. Roy. Astron. Soc.}

\begin{document}

\widetext
\leftline{Version 1 as of \today}
\leftline{Primary authors: Maayane T. Soumagnac}
\leftline{Accepted for publication in Physical Review Letters}


\title{Large scale distribution of total mass versus luminous matter from Baryon Acoustic
Oscillations:\\
First search in the SDSS-III BOSS Data Release 10}
\affiliation{Department of Particle Physics and Astrophysics, Weizmann Institute of Science, Rehovot 76100, Israel}
\affiliation{Raymond and Beverly Sackler School of Physics and Astronomy, Tel Aviv University, Tel Aviv 69978, Israel}
\affiliation{Institut d'Astrophysique de Paris, Institut Lagrange de Paris, CNRS, UPMC Univ Paris 06, UMR7095, 98 bis, boulevard Arago, F-75014 Paris,
France}
\affiliation{Department of Astrophysics, University of Oxford, Denys Wilkinson Building, Keble Road, Oxford OX1 3RH, UK}
\affiliation{Korea Astronomy and Space Science Institute, 776, Daedeokdae-ro, Yuseong-gu, Daejeon, 305-348, Korea}
\affiliation{Astronomy Department, Harvard University, 60 Garden Street, Cambridge, MA 02138, USA}
\affiliation{Center for Cosmology and Astro-Particle Physics, The Ohio State University, 191 West Woodruff Avenue, Columbus, OH 43210, USA}
\affiliation{Department of Physics and Astronomy, University College London, Gower Street, London WC1E6BT, UK}
\affiliation{Department of Physics and Electronics, Rhodes University, PO Box 94, Grahamstown, 6140 South Africa}
\author{M.T.~Soumagnac} \affiliation{Department of Particle Physics and Astrophysics, Weizmann Institute of Science, Rehovot 76100, Israel}
\author{R. Barkana} \affiliation{Raymond and Beverly Sackler School of Physics and Astronomy, Tel Aviv University, Tel Aviv 69978, Israel} \affiliation{Institut d'Astrophysique de Paris, Institut Lagrange de Paris, CNRS, UPMC Univ Paris 06, UMR7095, 98 bis, boulevard Arago, F-75014 Paris, France} \affiliation{Department of Astrophysics, University of Oxford, Denys Wilkinson Building, Keble Road, Oxford OX1 3RH, UK}
\author{C. G. Sabiu} \affiliation{Korea Astronomy and Space Science Institute, 776, Daedeokdae-ro, Yuseong-gu, Daejeon, 305-348, Korea}
\author{A. Loeb} \affiliation{Astronomy Department, Harvard University, 60 Garden Street, Cambridge, MA 02138, USA}
\author{A. J. Ross} \affiliation{Center for Cosmology and Astro-Particle Physics, The Ohio State University, 191 West Woodruff Avenue, Columbus, OH 43210, USA}
\author{F. B. Abdalla} \affiliation{Department of Physics and Astronomy, University College London, Gower Street, London WC1E6BT, UK}\affiliation{Department of Physics and Electronics, Rhodes University, PO Box 94, Grahamstown, 6140 South Africa}
\author{S. T. Balan} \affiliation{Department of Physics and Astronomy, University College London, Gower Street, London WC1E6BT, UK}
\author{O. Lahav}\affiliation{Department of Physics and Astronomy, University College London, Gower Street, London WC1E6BT, UK}
%

%
%
\vskip 0.25cm
\date{\today}

\begin{abstract}
  Baryon Acoustic Oscillations (BAOs) in the early Universe are
  predicted to leave an as yet undetected signature on the relative
  clustering of total mass versus luminous matter. A detection of this effect would
  provide an important confirmation of the standard cosmological
  paradigm and constrain alternatives to dark matter as well as
  non-standard fluctuations such as Compensated Isocurvature
  Perturbations (CIPs). We conduct the first observational search for
  this effect, by comparing the number-weighted and luminosity-weighted correlation functions, using the SDSS-III BOSS Data Release 10 CMASS sample.
  When including CIPs in our model, we formally obtain evidence at $3.2\sigma$ of the relative clustering signature and a limit that
  matches the existing upper limits on the amplitude of CIPs.
  However, various tests suggest that these results are not yet
  robust, perhaps due to systematic biases in the data. The method
  developed in this Letter, used with more accurate future data such
  as that from DESI, is likely to confirm or disprove our preliminary evidence.
\end{abstract}

\maketitle
{\it Introduction - } 
In the hot and dense early Universe, the interplay between the plasma
pressure and the radiation pressure resulted in ``sound waves'':
baryonic shells propagating around each initial overdensity of matter.
At the time of recombination, approximately 370,000 years after the
Big Bang, these baryonic sound waves froze, leaving an
  oscillatory signature in the distribution of baryons. After
  recombination, in the absence of significant radiation pressure, the
  distributions of baryons and Cold Dark Matter (CDM) grew
  increasingly similar due to their mutual gravitational attraction.
  This resulted in a bump in the two point correlation function of the
  positions of galaxies, a signature known as ``Baryon Acoustic
  Oscillations'' (BAOs). This feature has served, since its detection
  in the 2dF Galaxy Redshift Survey (2dFGRS) and the Sloan Digital Sky
  Survey (SDSS) \citep{Percival2001, Eisenstein2005, Cole2005}, as a
  precious cosmological tool to probe the expansion of the
  Universe. 

Another important aspect of BAOs, which has not yet been detected, is
a related imprint on the clustering of light {\it
    relative} to mass.  Indeed, while gravity helped the baryons catch
  up with the CDM distribution after recombination, this asymptotic
  process remains incomplete and the resulting scale dependence of
the ratio of baryonic to total matter contrasts, $\delta_{\rm
  b}/\delta_{\rm tot}$, should still be observable at present.
Detecting this scale dependence would offer a new angle to compare the
large scale distribution of light versus mass, an effort that dates
back to the 1980s \citep{Lahav1987, Erdogdu2006}.

Specifically, the detection of the scale dependence of $\delta_{\rm
  b}/\delta_{\rm tot}$ imprinted by BAOs is important for three
reasons: the {\it detection} of the effect would provide a direct
measurement of a difference in the large-scale clustering of mass and
light and thus a novel confirmation of the standard cosmological
paradigm (especially if the precise theoretically-predicted form of
the scale dependence is verified). It would present a strong challenge
to alternative theories of gravity, specifically non-dark matter
models such as MOND \citep{Milgrom1994} and its
  extensions \citep{Bekenstein2004} or Modified Gravity
  \citep{Moffat2006}. Direct evidence for the existence of dark matter
  includes the data from the bullet cluster \citep{Clowe2006}. The
  measurement of the scale dependence of $\delta_{\rm b}/\delta_{\rm
    tot}$ from BAOs, would provide evidence comparable to the bullet
  cluster, with the significant advantage that this effect happens on
  linear scales and thus may be easier to interpret
  \citep{Brownstein2007}. The {\it amplitude} of the effect would
probe a novel aspect of galaxy formation, specifically calibrating the
dependence of the average mass-to-light ratio of galaxies on the
baryon mass fraction of their large-scale environment.  Finally, we
show in this paper that such a detection would also constrain the
amplitude of Compensated Isocurvature Perturbations (CIPs).

The measurement of the scale dependence of $\delta_{\rm b}/\delta_{\rm
  tot}$ requires one to compare observable tracers of $\delta_{\rm
  tot}$ and $\delta_{\rm b}$. In this Letter, we follow and extend
\footnote{We note that \citet{Barkana2011} wrote that measuring the
  scale-dependent bias of galaxies would probe a novel aspect of
  galaxy formation; in this Letter we extend the importance of such a
  measurement by pointing out two additional, even more significant
  consequences.} the proposal by \citet{Barkana2011} (hereafter BL11),
i.e., we use the number density $\delta_{\rm n}$ of galaxies as a
tracer of the total matter density fluctuation $\delta_{\rm tot}$ and
the absolute luminosity density of galaxies $\delta_{\rm L}$ as a
tracer of the baryonic density fluctuation $\delta_{\rm b}$. The idea
is as follows: the number density fluctuations $\delta_{\rm n}$ are
driven by the underlying total matter density fluctuation $\delta_{\rm
  tot}$, with a bias (i.e., ratio) $b_{\rm n,t}$, which should be
approximately constant on large scales. On the other hand, an area
with a higher baryonic mass fraction $\delta_{\rm b}/\delta_{\rm tot}$
than average is expected to produce more stars per unit total mass,
hence more luminous matter and to result in galaxies with a lower
mass-to-light ratio. As a result, the luminosity-weighted density
fluctuation, $\delta_{\rm L}$, traces a combination of $\delta_{\rm
  tot}$ and $\delta_{\rm b}$.  Therefore, the scale dependence of
$\delta_{\rm b}/\delta_{\rm tot}$ induced by BAOs should translate
into a scale dependence of $\delta_{\rm L}/\delta_{\rm n}$.

{\it Predictions -} 
%
%
BL11 provide a model for the tracers $\delta_{\rm n}$
and $\delta_{\rm L}$ of the quantities of interest $\delta_{\rm b}$
and $\delta_{\rm tot}$:
\begin{eqnarray}\label{eq:finaldeltan}
\delta_{\rm n}=\left(b_{\rm n,t}+Cb_{\rm L,t}+
    Cb_{\rm L,\Delta}[r(k)-r_{\rm lss}]\right)\delta_{\rm tot}\;,
\end{eqnarray}
\begin{eqnarray}
\label{eq:finaldeltal}
\delta_{\rm L}=\left(b_{\rm n,t}+(1+D)b_{\rm L,t}+
(1+D)b_{\rm L,\Delta}[r(k)-r_{\rm lss}]\right)\delta_{\rm tot}\;.
\end{eqnarray}
Within this model, bias factors $b_{\rm n,t}$ and
  $b_{\rm L,t}$ reflect the dependency of the number density and mean
  luminosity fluctuations on the underlying matter density fluctuation
  \footnote{The mean luminosity of galaxies may depend on their
    environment through their merger rate history, which is correlated
    with the local matter density.}. The mean luminosity fluctuations
  are also affected separately by the baryon fluctuations because the
  luminosity depends on the gas fraction in haloes, which itself
  depends - through the non-linear process of halo collapse - on the
  baryon fraction of the surroundings.  The
  parameter $b_{\rm L,\Delta}$ quantifies the effect we search for: it
  is an effective bias factor that measures the overall dependence of
  galaxy luminosity on the underlying difference $\Delta$ between the
  baryon and total density fluctuations; $C$ and $D$ quantify effects
emerging in surveys where the observed sample is flux-limited (which
introduces additional dependences on galaxy luminosity); and $r(k)$ is
the fractional baryon deviation $r(k) = \delta_{\rm b}/\delta_{\rm
  tot} - 1$, which can be predicted from the initial power spectra,
and which approaches a constant (i.e., scale-independent though
redshift-dependent) value $r_{\rm lss}$ on scales below the BAOs.
Eqs.~(\ref{eq:finaldeltan}) and (\ref{eq:finaldeltal}) refer to
amplitudes at a given wavenumber $k$ of Fourier-decomposed fluctuation
fields.

{\it Compensated Isocurvature Perturbations - }
The measurement of the relation between dark matter and baryons is
related to the search for CIPs \citep{Grin2014}. Measurements of
primordial density perturbations are consistent with adiabatic initial
conditions, for which the ratios of neutrino, photon, baryon and CDM
energy densities are initially spatially constant. Indeed, the
simplest inflationary models predict adiabatic fluctuations
\citep{Guth1982,Linde1982}.  However, more complex inflationary
scenarios \citep{Brandenberger1994, Linde1984, Axenides1983} predict
fluctuations in the relative number densities of different species,
known as Isocurvature Perturbations.  Cosmic microwave background
(CMB) temperature anisotropies limit a matter versus radiation
isocurvature mode to a few percent of the adiabatic modes
\citep{Planck2015}. CIPs, however, are specifically perturbations in
the baryon density $\delta_{\rm b}$ that are compensated for by
corresponding fluctuations in the CDM $\delta_{\rm CDM}$ (so that the
total density is unchanged).

Such fluctuations are hard to detect, since gravity (and its effect on
everything from galaxy numbers to CMB fluctuations) only depends on
the total density. The uniformity of the baryon fraction of galaxy
clusters \citep{Holder2010} gives an upper limit on CIPs corresponding
to $\Delta_{\rm cl} < 7.7\%$, where $\Delta_{\rm cl}$ is the RMS
fluctuation in the baryon to CDM density ratio on galaxy cluster
scales. Non-linear effects on the CMB give a similar current limit,
$\Delta_{\rm cl} < 11\%$ \citep{Grin2014}. These constraints may be
improved with future cosmological 21-cm absorption observations
\citep{Gordon2009}. In this paper we added possible CIPs to the BL11
model under the standard assumption of a scale-invariant power
spectrum for this field.

{\it Model in terms of correlation function - }
The observable quantities in galaxy surveys are not the fluctuations
$\delta_{\rm n}$ and $\delta_{\rm L}$ but rather the two point
statistics of such tracers, namely the power spectrum or the two-point
correlation function (2PCF). We reformulate the observational proposal
of BL11 in terms of the 2PCF, defined as
\begin{eqnarray} \label{eq:xilin}
\xi({\bf x},{\bf y}) \equiv \frac{1}{2\pi ^2}\int k^2P(k)j_0(ks)dk \;,
\end{eqnarray}
where $s=|{\bf x} - {\bf y}|$ and $P(k)$ is the matter power spectrum
defined by $\left<\delta(\bf k)\delta(\bf k')\right>\equiv
P(k)\delta^D({\bf k}-{\bf k'})$. Following the notation of BL11, we find that the
observable 2PCFs $\xi_{\rm n}$ (of the galaxy number density) and
$\xi_{\rm L}$ (of the galaxy luminosity density) can be expressed with three
theoretically-predicted functions, $\xi_{\rm tot}$, $\xi_{\rm add}$,
and $\xi_{\rm CIP}$, the set of five BL11 parameters from
eqs.~(\ref{eq:finaldeltan}) and (\ref{eq:finaldeltal}) and the
parameter $B_{\rm CIP}$ (which determines the amplitude
of CIPs).
Defining total effective bias parameters $ B_{\rm n,t}=b_{\rm
  n,t}+Cb_{\rm L,t}$, $B_{\rm n,\Delta}=Cb_{\rm L,\Delta}$, $B_{\rm
  L,t}=b_{\rm n,t}+(1+D)b_{\rm L,t}$, and $B_{\rm
  L,\Delta}=(1+D)b_{\rm L,\Delta}$, our model equations are:
\begin{eqnarray}\label{eq:xin_new}
  \xi_{\rm n} =B_{\rm n,t}^2\cdot\xi_{\rm tot}+ 
2B_{\rm n,t}B_{\rm n,\Delta}\cdot\xi_{\rm add}+B_{\rm n,\Delta}^2B_{\rm CIP}\cdot\hat{\xi}_{\rm CIP}\;,\\
\label{eq:xil}
\xi_{\rm L} =B_{\rm L,t}^2\cdot\xi_{\rm tot}+ 
2B_{\rm L,t}B_{\rm L,\Delta}\cdot\xi_{\rm add}+
B_{\rm L,\Delta}^2B_{\rm CIP}\cdot \hat{\xi}_{\rm CIP}\;,
\end{eqnarray}
where (unlike the other $\xi$ terms) we have separated $\xi_{\rm CIP}$
into its shape $\hat{\xi}_{\rm CIP}$ and its amplitude $B_{\rm CIP}$.
In order to model the correlation functions, we begin with linear
perturbation theory, for which $\xi_{\rm tot}(s)$ is given 
by Eq.~\ref{eq:xilin},
\[
\xi_{\rm add}(s)=\frac{1}{2\pi ^2} \int k^2[r(k)-
r_{\rm lss}]P(k) j_0(ks)dk \;,
\]
\[\xi_{\rm CIP}(s)\equiv B_{\rm CIP}\cdot \hat{\xi}_{\rm CIP}(s)=
\frac{B_{\rm CIP}}{2\pi ^2} \int \frac{j_0(ks)}{k}dk\; .\] Our full
model with the addition of corrections for non-linear clustering and
for systematic effects is presented in the Supplemental Material \footnote{See Supplemental Material at [URL will be inserted by publisher] for our full
model with the addition of corrections for non-linear clustering and
for systematic effects.}.

{\it Measurement - } In all this analysis, we use the latest public
data release from the SDSS-III Baryon Oscillation Spectroscopic Survey
(BOSS), DR10 \footnote{http://www.sdss3.org/dr10/}
\citep{Ahn2014,Anderson2014}. The BOSS collaboration has analysed a
larger set of data, denoted DR11 in \cite{Anderson2014}, which will be
publicly released with the final BOSS data set. For
  both DR10 and DR11, the BOSS collaboration has made public some
  ``final products'', namely their measurement of $\xi_n$ and the
  associated covariance matrix (but not $\xi_L$), and we checked that
  they are in good agreement with our measurement of $\xi_n$ and give
  a reasonable fit to the $\xi_n$ part of our model. Several
practical problems inhibit our ability to accurately measure the 2PCF
of the galaxy distribution. The discreet sampling by individual
galaxies of the smooth density field leads to shot noise on small
scales. Other difficulties arise from the irregular shape of galaxy
surveys in angular sky coverage, due to dust extinction, bright stars,
tracking of the telescope,
etc. 
In this work, the two-point correlation functions $\xi_{\rm n}$ and
$\xi_{\rm L}$, are computed using the optimal Landy-Szalay estimator
\citep{Landy1993} which requires the creation of a catalog of random
positions.

\begin{figure}
\includegraphics[scale=0.4]{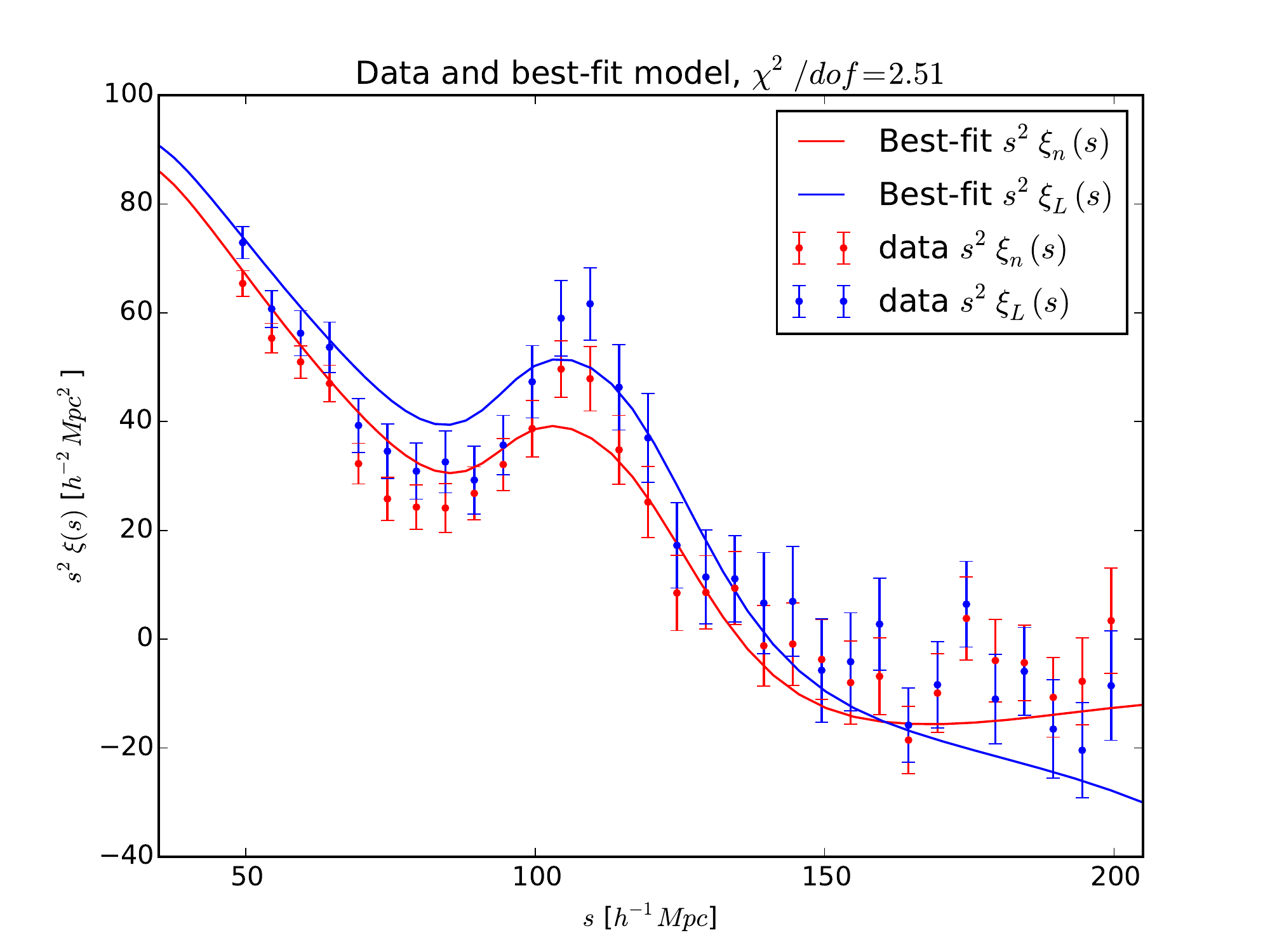}
\caption{Our measurement of $\xi_{\rm L}$ (blue) and $\xi_{\rm n}$
  (red) [times $s^2$], using $31$ radial bins, and our best-fit
  maximum likelihood model (allowing all parameters to be non-zero).
  The best fit corresponds to $\chi^2/\rm{dof}=2.51$
    , where $\rm{dof}$ is the number of degrees of freedom in the fit.
    This high value of $\chi^2/\rm{dof}$ is partially due to the
    highly correlated errors among the various binned measurements and
    perhaps systematic errors (it drops to $\sim1.5$ when using $21$
    bins), which also make the fits difficult to judge visually.}
\label{fig:xi}
\end{figure}

We calculate the two-point correlation function $\xi_{\rm L}$ of the
absolute luminosity density fluctuations using the same estimator and
algorithms as for $\xi_{\rm n}$, but weighting each object with its
absolute luminosity. The absolute luminosity is calculated using the
$i$-band photometric data, from the CMASS DR10 catalogs.
We use a Jackknife (JK) resampling technique, as in
\citet{Scranton2002}, to compute the full covariance matrix for the
joint measurement of $\xi_{\rm n}(r)$ and $\xi_{\rm L}(r)$. This
technique differs from the method adopted by the BOSS collaboration,
where 600 mock catalogs were produced and used to estimate the
covariance matrix for the fit \citep{Manera2013,Percival2014}. 
Figure~\ref{fig:xi} shows our measurement of $\xi_{\rm L}$ and
$\xi_{\rm n}$ and our best-fit model, as detailed in the next section.

{\it Model Fitting - }
%
We adopt the model-fitting formalism of of \citet{Hogg2010}, and
assume that the only source for deviation of our data points from the
model described by equations~\ref{eq:xin_new} and~\ref{eq:xil} is an
offset in the $\xi$ direction, drawn from a Gaussian distribution of
zero mean and known covariances. We wish to get the set of parameters
$\theta$ that maximizes the probability of our model $\mathcal{M}$
given the data $\mathcal{D}$, i.e., the posterior probability
distribution ${\rm Pr}(\theta|\{\mathcal{D},\mathcal{M}\})$.
%
We make a conservative choice of uniform (not ``informative'') priors
for the parameters of our model: The prior on $B_{\rm L,\Delta} \in
[-10,10]$ is intentionally taken to be broad, although BL11 forecasted
it to be around 2.6.  The best current limits on $\Delta_{\rm cl}$
correspond \citep{Grin2014} to an upper limit of $B_{\rm CIP} \approx
5 \times 10^{-3}$ from clusters or $1.1 \times 10^{-2}$ independently
from the CMB; we allowed a much broader range and applied the prior
$B_{\rm CIP} \in [-0.3,0.3]$. The other priors are given in the
Supplemental Material \footnote{See Supplemental Material at [URL will be inserted by publisher] for the priors on the other parameters.}.

\begin{figure}
\includegraphics[scale=0.45]{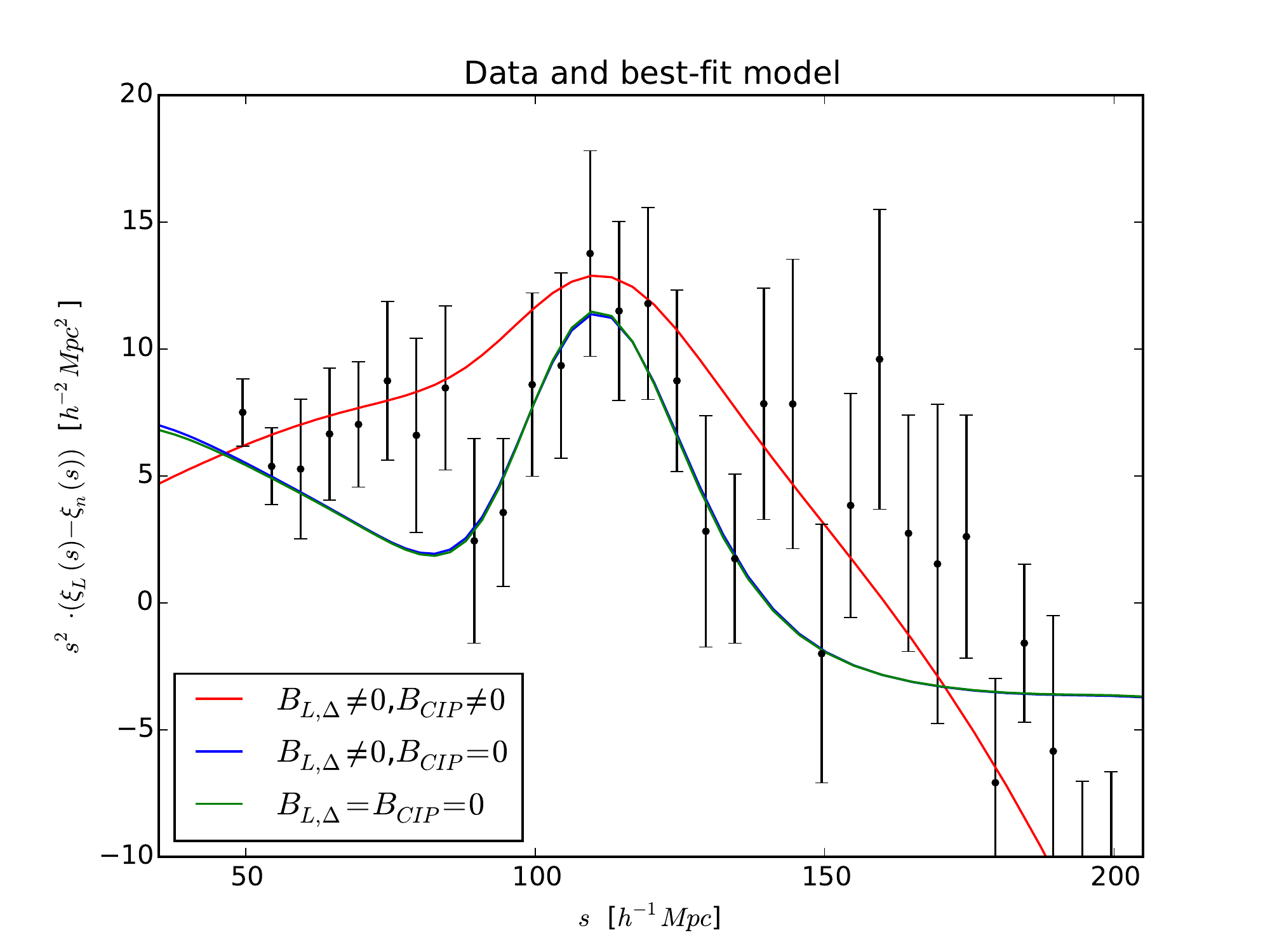}
\caption{Our measurement of the difference $\xi_{\rm L}-\xi_{\rm n}$
  (times $s^2$), using $31$ radial bins, and the same quantity in our
  best-fit model.  The red line corresponds to our full model, the
  blue line corresponds to a model with $B_{\rm CIP}=0$, and the green
  line corresponds to a model with $B_{\rm CIP}=B_{\rm L,\Delta}=0$.}
\label{fig:diff}
\end{figure}

In the case of a non-informative prior, the optimisation of the
likelihood function corresponds to the maximum of the posterior probability distribution, i.e.,
the maximum a posteriori value. To estimate the uncertainty in the
maximum a posteriori value of each parameter, we obtain the
distribution of parameters that is consistent with our data, and
marginalise over it to get the distribution of each parameter. We did
this using the Monte Carlo Markov Chain (MCMC) algorithm {\it MultiNest} \citep{Feroz2008,Skilling2009},
to sample from the posterior probability
distribution, and quote 
$1\sigma$ limits. 
We consider two cases, corresponding to the presence or absence of
CIPs. 
In Figures~\ref{fig:xi} and~\ref{fig:diff}, we show the data and best
fits for the correlation functions $r^2\xi_{\rm n}$ and $r^2\xi_{\rm
  L}$, and for a key quantity, their difference $r^2(\xi_{\rm
  L}-\xi_{\rm n})$. We checked that all the following conclusions are
not significantly altered when adding k-corrections
  and evolutionary corrections and when simulating the effect of the
photometric errors on the measurement of $\xi_{\rm L}$.

{\it Results -}
When we allow CIPS, i.e., $B_{\rm CIP}\neq0$, we obtain evidence at $3.2\sigma$ of $B_{\rm L,\Delta}>0.4$ (and evidence that $|B_{\rm L,\Delta}|>0.4$ at $3.7\sigma$),
which indicates the presence of the effect we search for, that of the
baryon-CDM difference on galaxy luminosity. Moreover,
  the $1\sigma$ range of $1.1<B_{\rm L,\Delta}<2.8$ is consistent with
  the prediction of BL11 of $B_{\rm L,\Delta} \approx 2.6$ (our
  maximum likelihood value is $3.9$) \footnote{In BL11, the $2.6$
    value is predicted along with the expectations of $B_{\rm
      n,\Delta} \approx 0$, and $B_{\rm n,t}$ and $B_{\rm L,t}$
    approximately equal.}. In addition, our best-fit value of $B_{\rm
    CIP}$ is $2.3\times 10^{-3}$, with a $2\sigma$ upper limit
  of 
  $B_{\rm CIP}=6.4\times 10^{-2}$, which is within an order of
  magnitude of the best existing limits noted previously. A full
  tabulation of our best-fit parameters, plus results with a smaller
  number of data bins, are given in the Supplemental Material \footnote{See Supplemental Material at [URL will be inserted by publisher] for a full
  tabulation of our best-fit parameters, plus results with a smaller
  number of data bins.}.

To determine whether we detect a scale-dependent bias of the
luminosity correlation function requires answering the following
question: do the data support the inclusion of a non-zero extra
parameter $B_{\rm L,\Delta}$? Rather than a question of parameter
estimation, this is a question of model comparison between two models
$\mathcal{M}$, with or without $B_{\rm
  L,\Delta}$. 
Within a Bayesian framework \cite{Verde2013}, the key quantity for
comparing them is the Evidence (or model-averaged likelihood), $E=\int
Pr(\theta|\mathcal{M})Pr(\mathcal{D}|\theta,\mathcal{M})d\theta$. The
ratio of the evidences, also called the Bayes factor, can be
calculated using the multimodal nested sampling algorithm, {\it
  MultiNest}
\citep{Feroz2008}. 
%
In the $B_{\rm CIP}\neq0$ case, the evidence ratio is $\ln(E_{B_{\rm
    L,\Delta}\neq0}/E_{B_{\rm L,\Delta}=0})= 6.08\pm0.23$,
which we interpret as strong evidence for $B_{\rm
  L,\Delta}\neq0$ 
according to the slightly modified Jeffreys' scale \citep{Jeffrey1961,
  Kass1995,Verde2013}. 

However, we believe that the results are not yet robust enough for
making strong claims. For one thing, if we model the data without
allowing for CIPs (i.e., setting $B_{\rm CIP} = 0$), the evidence for
a detection of non-zero $B_{\rm L,\Delta}$ goes away. Our $1\sigma$ range of $-1.0<B_{\rm L,\Delta}<7.8$ in
  that case is consistent with the previous ($B_{\rm CIP} \ne 0$)
case and with the BL11 prediction, but also with a value of zero. This
lack of evidence is reflected by the evidence ratio $\ln(E_{B_{\rm L,\Delta}\neq0}/E_{B_{\rm
      L,\Delta}=0})=0\pm0.23$
, corresponding to no evidence toward one model versus the other
\footnote{When setting $B_{\rm CIP}=0$ or $B_{\rm
      CIP}=B_{\rm L,\Delta}=0$, we obtain
    $\chi^2/\rm{dof}=2.74$. 
    This reflects the fact that there is little difference between the
    case $B_{\rm CIP}=0$ and the case $B_{\rm CIP}=B_{\rm
      L,\Delta}=0$, which is consistent with the lack of evidence
    revealed by the evidence ratios.}. The high value
  of $\chi^2/\rm{dof}$, partially due to the high correlated errors
  between the various binned measurements \footnote{When using $21$
    bins, the value of $\chi^2/\rm{dof}$ decreases to $\sim 1.5$.}
  points at the need to eliminate systematic errors or try more
  sophisticated models in future implementations of this method. The
  fact that the parameter values are affected by the choice of the
  number of radial bins is another sign of the lack of robustness of
  our result. More generally, disentangling the various effects is
difficult, since the model of equations~\ref{eq:xin_new}
and~\ref{eq:xil} shows that any ability to set a limit on CIPs depends
on a definitive detection of non-zero $B_{\rm L,\Delta}$ (and/or
$B_{\rm n,\Delta}$).  Conversely, the presence of a significant CIP
term in the fit strongly affects the best-fit values of $B_{\rm
  L,\Delta}$ and $B_{\rm n,\Delta}$. Trying to measure two novel
effects (one of them expected but with an uncertain amplitude, the
other highly speculative) when they are entangled in this way is
tricky. Another difficulty comes from the fact that $\hat{\xi}_{\rm
  CIP}$ has a smooth shape (in contrast with BAO-scale features in
$\xi_{\rm tot}$ and $\xi_{\rm add}$), and such a slowly-varying term
may more easily be emulated by systematic effects; we note that
standard BAO measurements (e.g., \citep{Percival2014}) typically add
several such ``nuisance'' terms, which are necessary to get good fits
to the data, do not significantly affect the BAO peak/trough
positions, but are not theoretically well-understood. We also note that several of our best-fit parameters
  change strongly between the zero and non-zero $B_{\rm CIP}$.
  Especially worrying is that in our full model, a strongly negative
  $B_{\rm sys,L}$ makes a large negative contribution that is nearly
  canceled out by large positive contributions from the other terms.


{\it Conclusion - }
We have compared the large-scale distribution of total
  mass and luminous matter, through measurement of the
number-weighted and luminosity-weighted galaxy correlation functions
$\xi_{\rm n}$ and $\xi_{\rm L}$ in the latest public data release from
the SDSS-III Baryon Oscillation Spectroscopic Survey (BOSS). We have
shown that such a measurement is potentially of great importance for
verifying the standard cosmological model and for putting new limits
on non-standard possibilities. In particular, such a measurement can
be used to detect the large-scale modulation from BAOs of the ratio of
baryonic matter to total matter. Within the framework of the model of
\citet{Barkana2011}, the effect of this modulation on galaxy surveys
is characterised by a parameter, $B_{\rm L,\Delta}$, which we have
measured in the BOSS CMASS DR10 data. When including non-standard (but
currently weakly constrained) CIPs in our model, we obtain evidence at $3.2\sigma$ of the modulation effect with
a value of $B_{\rm L,\Delta}$ consistent with the theoretical
prediction, and an upper limit on the CIP amplitude that is within an
order of magnitude of the best existing limits.  However, current data limit the robustness of this test and we believe our results only demonstrate that current data
are on the threshold of detecting the BAO-induced modulation and
setting strong limits on
CIPs. 
Future observational efforts, such as the Dark Energy Spectroscopic
Instrument (DESI) \citep{Levi2013}, will provide more accurate data. In particular, while we used $\sim 0.5M$ galaxies for
  this analysis, DESI will have $\sim20M$ galaxies, which will reduce the
  statistical error on the correlation function measurement and
  increase the redshift coverage. The better quality imaging will
  reduce the error on the luminosity measurement and subsequently on
  $\xi_L$. We expect new data sets, as well as more
  robust theoretical modeling, to improve the robustness of the evidence, and thus to definitively verify or rule out
the predicted
effect. 

\section{Acknowledgments}
%

MTS would like to thank support from the University College London
Perren and Impact studenships, from the Benoziyo Center for Astrophysics and the Department of 
Particle Physics and Astrophysics at the Weizmann Institute of Science and from the department of Astronomy and Astrophysics
of Tel Aviv University. The authors thank Korea Institute for
Advanced Study for providing computing resources (KIAS Center for
Advanced Computation Linux Cluster) for this work. R.B.\ acknowledges
Israel Science Foundation grant 823/09, the Ministry of Science and
Technology, Israel, and a Leverhulme Trust Visiting
Professorship. R.B.'s work has also been done within the Labex
Institut Lagrange de Paris (ILP, reference ANR-10-LABX-63) part of the
Idex SUPER, and received financial state aid managed by the Agence
Nationale de la Recherche, as part of the programme Investissements
d'avenir under the reference ANR-11-IDEX-0004-02. AL was supported in
part by NSF grant AST-1312034. FBA acknowledges the support of the
Royal Society via a University Research Fellowship. AJR is thankful
for support from University of Portsmouth Research Infrastructure
Funding. OL acknowledges a Royal Society Wolfson Research Merit Award,
a Leverhulme Senior Research Fellowship and an Advanced Grant from the
European Research Council. STB's research at UCL is funded through an
Advanced Grant from the European Research Council awarded to OL.

\end{document}